\documentclass[a4paper, fleqn, usenatbib, useAMS]{mnras}

\def\aap{AA}
\def\apjl{ApJL}
\def\apjs{ApJS}
\def\mnras{MNRAS}
\def\apj{ApJ}

\def\araa{ARAA}
\usepackage{multicol}        
\usepackage{graphicx}
\usepackage{float}
\usepackage{bm} 
\usepackage{amssymb}
\usepackage{hyperref}
\usepackage{amsmath} 
\usepackage{pdflscape}
\usepackage[export]{adjustbox}
\usepackage{mathrsfs} 
\usepackage{mathtools}

\title[The accreted stellar halo revealing halo assembly]{The accreted stellar halo as a window on halo assembly in L$^*$ galaxies}
\author[Nicola C. Amorisco]{Nicola C. Amorisco$^{1, 2}$\thanks{E-mail:
nicola.amorisco@cfa.harvard.edu} \\
$^{1}$Max Planck Institute for Astrophysics,  Karl-Schwarzschild-Strasse 1,  D-85740 Garching,  Germany, \\
$^{2}$Institute for Theory and Computation,  Harvard-Smithsonian Center for Astrophysics,  60 Garden St.0,  MS-51,  Cambridge,  MA 02138,  USA}

\begin{document}



\maketitle

\label{firstpage}

\begin{abstract}

Theory and observations agree that the accreted stellar halos (ASHs) of Milky Way-like galaxies display significant scatter. 
I take advantage of this stochasticity to invert the link between halo assembly history (HAH) and ASH, using mock ASHs corresponding to 750 $\Lambda$CDM HAHs, sharing a final virial mass {of $M_{h}(z=0)=10^{12.25}M_\odot$}. 
Hosts with poor/rich ASHs assemble following orthogonal growth-patterns. Hosts with rich ASHs experience accretion events (AEs) with high virial mass ratios (HVMRs, $M_s/M_h\gtrsim 0.1$) at $0.5\lesssim z_{infall}\lesssim1.5$, in a phase of fast growth. This maximizes the accreted stellar mass under the condition these satellites are disrupted by $z=0$. At similar times, hosts with poor ASHs grow slowly through minor mergers, with only very recent HVMR AEs: {this results in a globally more abundant satellite population and in distinctive surviving massive satellites} (stellar mass $\log M_{s,*}/M_\odot\gtrsim 9$). Several properties of the Milky Way are in agreement with the predictions of this framework for hosts with poor, concentrated ASHs, including: i) the recent infall of Sagittarius and Magellanic Clouds, ii) the likely higher-than-average concentration of its dark halo, iii) the signatures of fast chemical enrichment of a sizable fraction of its halo stellar populations.

\end{abstract}


\section{Introduction} 

The accreted stellar halo (ASH) of a galaxy collects all those stars
that were born {\it ex-situ}, within another less massive galaxy, and 
that accumulated around their current host through hierarchical 
merging \citep[e.g.][]{Egg62,SZ78,SW91,KJ08}. As such, the ASH 
represents a record of the assembly process of galaxies through 
cosmic time, by means of which it is possible, at least in theory, to 
test the prevailing cosmological paradigm.

Currently, however, a clear bridge between halo
assembly history (HAH) and the properties of the ASH is missing, 
and the inversion of this connection appears difficult, despite significant effort
\citep[e.g.][and references therein]{BJ05,AC10,AD14,AP14}. 
A fundamental reason for this is that the properties of the ASH
are a function of an extremely large number of free parameters: at least a handful
are needed to determine how each single `building block' deposits 
its stars, even under the crudest simplifications \citep{NA17a}.
This needs to be factored by the number of contributing satellite galaxies, 
and by the stochasticity of merging histories. As a result, it remains difficult 
and possibly misleading to draw conclusions based on small samples of 
simulations, regardless of their realism. Here, I use tens of thousands 
of toy models \citep[][]{NA16,NA17b} to systematically compare the HAHs of 
$L^*$ galaxies differing in their ASHs.

It is interesting to note that, within a $\Lambda$CDM universe, the link 
between HAH and ASH can be expected to become less clear with 
increasing host halo mass. $\Lambda$CDM mean merging histories are
approximately independent of host mass  \citep[e.g.,][]{GW08,Fa10}, implying very similar numbers 
of minor mergers per virial mass ratio $M_s/M_h$, {where $M_s$ and $M_h$ are respectively the
virial masses of satellite and host, at the redshift of infall}. In turn, 
the efficiency of haloes at forming galaxies is instead a strong function of halo 
mass \citep[e.g.,][]{BM10,GW10}.
As a consequence:
\begin{itemize}
\item{for Milky Way-like (MW) hosts, 
the break in the stellar-to-halo mass relation (SHMR) and its steepness
below the break \citep[e.g.,][]{PB13,GK14}, ensure that accretion events (AEs) with high 
virial mass ratios (HVMRs) dominate the budget of accreted stars \citep[e.g.,][]{AC10, NA16, AD16}, 
leaving their distinctive fingerprints on both global and local properties of the ASH;}
\item{in massive ellipticals, instead, the numerous minor mergers contribute sufficient 
stellar material that ASHs better `converge' towards similar properties. 
This has an analogue in dark matter haloes themselves, where convergence is complete and a 
`universal profile' emerges \citep[e.g.,][]{NFW97,SW98,AL13}. }
\end{itemize}
Results of both theoretical analyses \citep[e.g.,][]{Pu07,AC13,AP14,RG16} 
and recent observations \citep[e.g.,][and references therein]{AMe16,Ha16,AMo16,KG12,RI14,BS11,AD14} 
are consistent with this interpretation, by showing that the scatter in the properties 
of the ASH increases in proceeding from massive hosts down to MW-like galaxies.

With an average of only 2.9 AEs at $z<3$ having a VMR $M_s/M_h > 0.1$ \citep{Fa10}, 
the properties of the ASH of a MW-like host are therefore dominated by 
Poisson noise in the galaxy's HAH, i.e. by whether, when, and how many of
these HVMR AEs actually took place.
In this Letter, I show that this stochasticity provides us with the opportunity of 
inverting the connection between HAH and ASH, using both global and 
local properties of the ASH as a window onto halo assembly (HA). 
Sect.~2 briefly introduces the models I employ, 
Sect.~3 illustrates results, Sect.~4 lays out the Conclusions.

\begin{figure}
\centering
\includegraphics[width=.78\columnwidth]{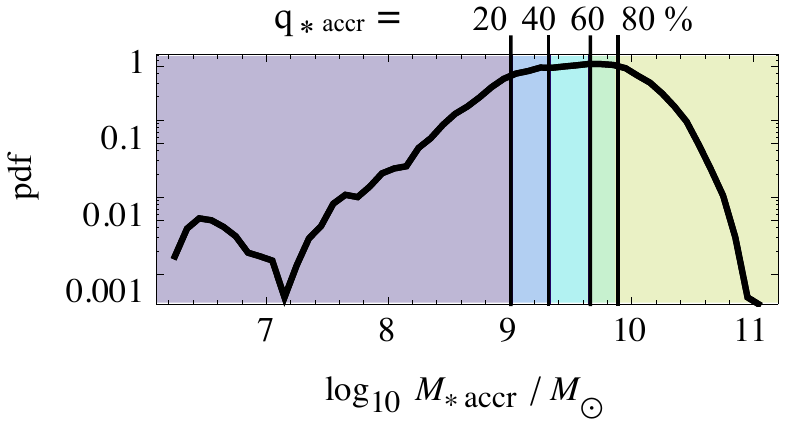}
\caption{The probability distribution function for the total mass in the 
accreted stellar halo, $M_{* accr}$. 750 individual halo assembly 
histories, sharing $M_{h}(z\!=\!0)=10^{12.25}M_\odot$, are used to
construct tens of thousands mock accreted haloes, to sample the scatter in the 
satellites' stellar-to-halo mass relation. Vertical lines separate the different quintiles.}
\end{figure}

\section{From halo assembly history to the accreted stellar halo}

In \citet{NA17a} I have shown that, under a set of assumptions that is essentially equivalent 
to the one adopted by particle-tagging techniques \citep[e.g.,][]{JB01,AC10,AC13,AC16},
the process of stellar deposition by a satellite galaxy can be reduced to a 
handful of dimensionless free parameters. This provides a strategy \citep{NA16}
to efficiently construct large numbers of mock ASHs, using a library of isolated, {dark matter only}
minor merger simulations. {In each minor merger, {both host and satellite 
have spherically symmetric NFW density profiles} and the satellite's stars are 
represented by its most bound particles, so that no additional assumptions have to be 
made regarding stripping}. 
Given a $\Lambda$CDM HAH  {and a standard SHMR for the satellites}, 
a toy ASH is built up like in a game of Lego, 
by adding up the (properly timed and rescaled) contributions of each accreted 
satellite, promptly recovered from the library. I refer to \citet{NA17b}
for details on this technique, and for a description of the properties of these mocks. 
There, I use a library of over 110 minor-merger simulations to construct three sets of ASHs,
for hosts with final virial mass of $\log M_h(z\!=\!0)/M_\odot \in \{11.8, 12.25,12.6\}$.
Each set explores 750 individual $\Lambda$CDM HAHs, each of which is used to generate 
tens of ASHs, so to sample the scatter in the SHMR.

Here, I employ the set of ASHs for the hosts with $\log M_h(z\!=\!0)/M_\odot = 12.25$. 
The probability density distribution for the total accreted stellar mass in the ASH, $M_{\rm * accr}$, 
is displayed in Fig.~1.  $M_{\rm * accr}$ collects debris deposited at all radii, and 
contributed by any satellite with VMR $\log M_s/M_h >-2.5$ {(defined at infall)}, 
at $z<4$.
Satellites can be either fully disrupted by $z=0$, or partially surviving in a bound 
remnant, in which case only the stripped stars contribute to $M_{\rm * accr}$. 
Fig.~1 shows that $M_{\rm * accr}$ ranges in the interval $6.2\lesssim \log M_{\rm * accr}/M_\odot\lesssim 11.1$,
confirming the extremely wide scatter in the global properties of the ASHs of MW-like galaxies.
While this is in very good quantitative agreement with the results of \citet[][]{AC13},
{the vast set of HAHs explored here allows me to first probe the surprising extent of the 
low-mass tail of the distribution, populated by ASHs that are under-massive by $>2$~dex with respect to 
the median} \citep[see also][]{NA17b}.

\begin{figure}
\centering
\includegraphics[width=.85\columnwidth]{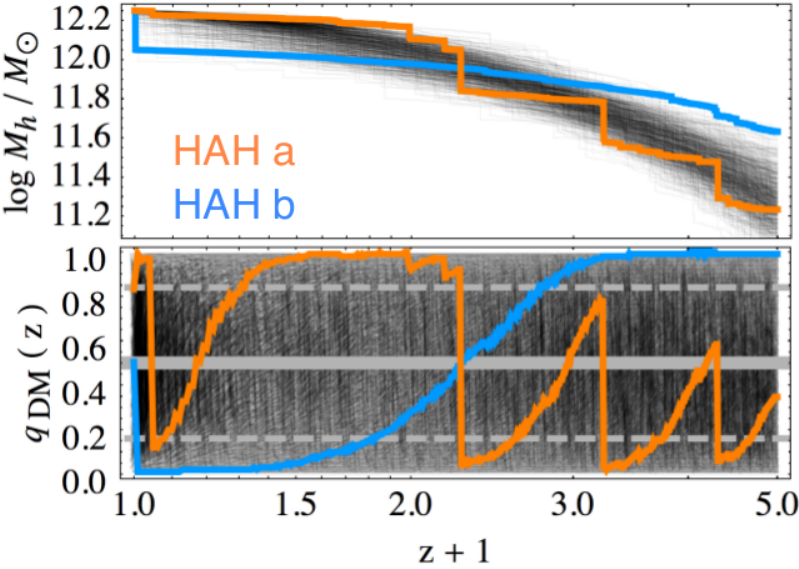}
\caption{The 750 studied HAHs, in terms of halo mass $M_{h}(z)$ (top panel), and redshift 
dependent quantile $q_{DM}(z)$ (bottom panel). The latter quantifies how the mass of each 
halo compares with the full ensemble at that redshift. Two different HAHs are highlighted.}
\end{figure}

The top panel of Fig.~2 illustrates the full set of 750 HAHs. 
The scatter increases towards higher redshifts, with haloes following different assembly patterns. 
Two different cases are highlighted: {\it HAH a} grows substantially at 
$z\gtrsim1$, by which time it is significantly more massive than average;
{\it HAH b}, instead, starts by being uncommonly massive at high redshift, 
and grows slowly thereafter, with a single HVMR AE, very recently. 
The bottom panel of Fig.~2 presents an alternative way of comparing 
HAHs: each individual track quantifies how the mass of each 
halo compares with the full ensemble, using the redshift dependent quantile $q_{DM}(z)$. 
By construction, haloes that are more (less) massive than the median at redshift $z$ have 
$q_{DM}(z) > (<)\ 0.5$ (grey solid line, with 16 and 84\% quantiles, dashed).
$q_{DM}$ readily visualises different modes of growth:
{\it HAH a} is characterised by 3 HVMR AEs at $z>1$, standing out as 
vertical `jumps'; {\it HAH b} is among the most massive 
haloes at high redshift, and then follows a monotonically decreasing track in $q_{DM}$, 
which stands for a slower-than-average growth sustained by minor mergers alone.

\begin{figure}
\centering
\includegraphics[width=\columnwidth]{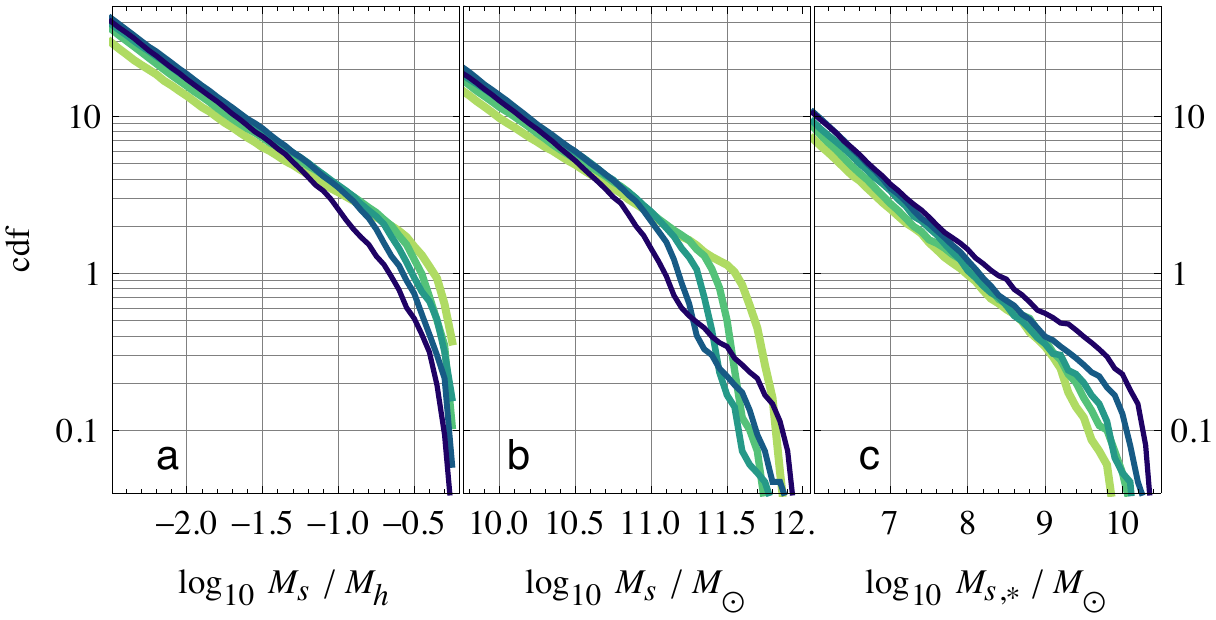}
\caption{Mean cumulative distribution functions for all accretion events 
{with virial mass ratio $M_s/M_h$ (panel a) or virial satellite mass $M_s$ (panel b)}.
Mean cumulative distribution functions for all surviving satellites with stellar mass $M_{s,*}$ at $z=0$.
Colors identify the different quintiles in $M_{* accr}$, as in Fig.~1.}
\end{figure}

\section{The accreted stellar halo shaping halo assembly}

\begin{figure*}
\centering
\includegraphics[width=\textwidth]{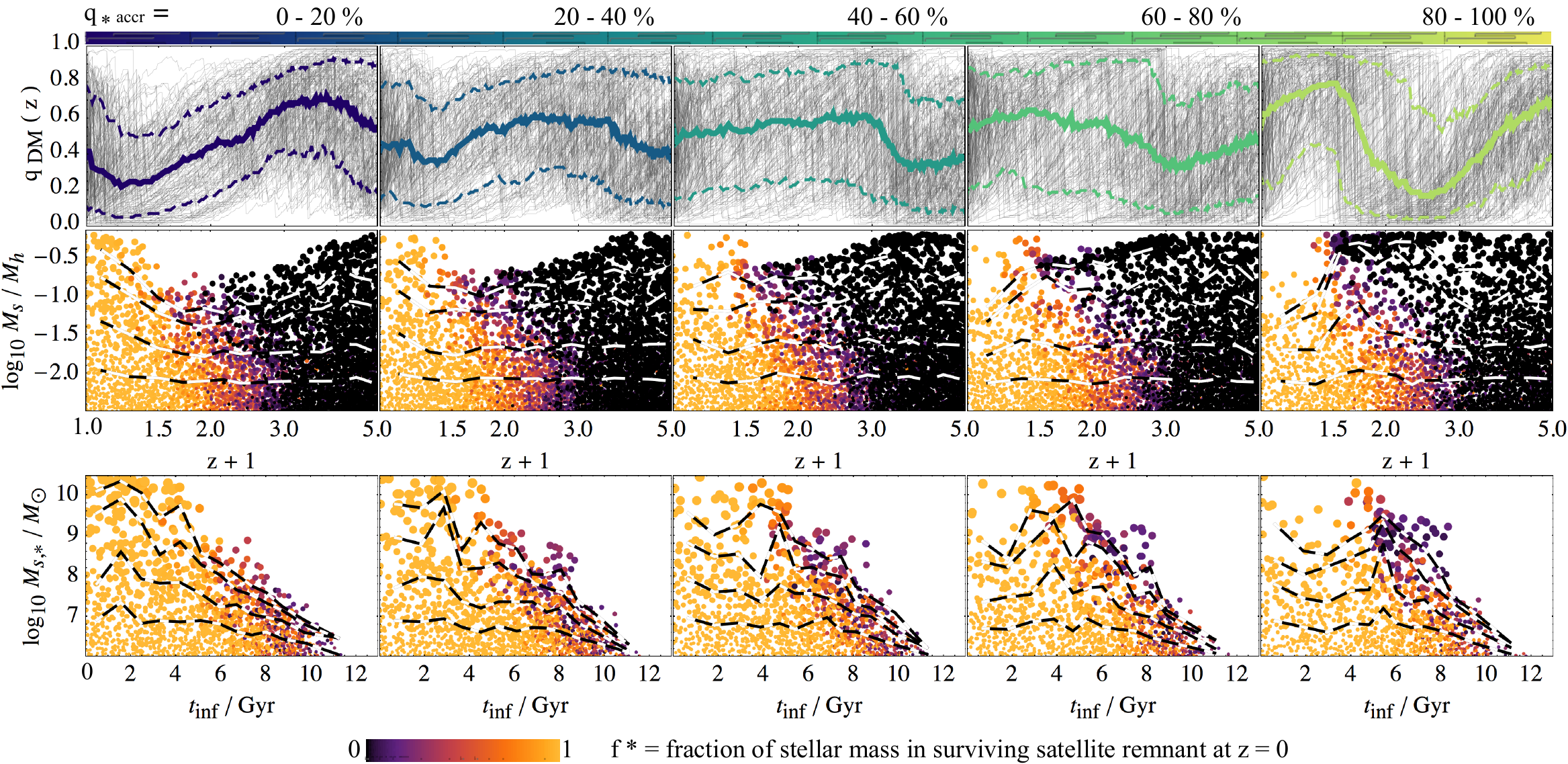}
\caption{
Columns identify different quintiles in total accreted stellar mass, increasing towards the right.
Top row: HAHs in terms of the quantile $q_{DM}(z)$; thin grey lines identify individual haloes, colored  
lines show the 16, 50 and 84\% quantiles (color-coding as in Figs.~1 and~2). 
Middle row: infall redshift and virial mass ratio $M_s  / M_h$ for all AEs; 
color-coding represents the fraction of stellar mass in the surviving satellite remnant at $z=0$, $f^*$; 
white-black dashed lines identify the 50, 75, 90, 95$\%$ quantiles as a function of infall time.
Bottom-row: infall time and stellar mass $M_{s,*}$ at $z=0$ for satellite remnants.
Color-coding and dashed lines as in the middle row.}
\end{figure*}

\subsection{Disentangling HAHs by total accreted stellar mass}

I partition the set of mock ASHs in quintiles, $q_{\rm * accr}$,
based on the total accreted stellar mass, as shown in Fig.~1. 
Each quintile collects 150 individual HAHs, resulting in similarly poor/rich ASHs.
{Values of the accreted stellar mass at the boundary between the different
families are $\log M_{*, accr}/M_{\odot}=\{9.02, 9.33, 9.65, 9.89\}$}.
Fig.~3 shows the mean cumulative distribution of AEs experienced by hosts within
these 5 families (color-coding as in Fig.~1), {in terms of the VMR $M_s/M_h$ 
at the time of infall in {\it panel a} and in terms of the satellite virial mass $M_s$
in {\it panel b}}. There is a clear ordering in the mean number of HVMR AEs, $M_s/M_h\gtrsim 0.1$,
or alternatively with $M_s\gtrsim 10.5$: 
hosts with richer ASHs collect the debris of a 
systematically larger number of massive building blocks.  
This is reversed at lower VMRs and satellite masses: where more dark matter
has been contributed by massive satellites there is systematically less room 
for low mass AEs.

This is mirrored in the current satellite population of each
family of hosts: {\it panel c} in Fig.~3 shows the mean cumulative distributions
of surviving satellites, in terms of their stellar mass $M_{s,*}$ at $z\!=\!0$. 
Hosts with rich ASHs feature a globally less numerous population of satellites, 
in line with {\it panels a} and {\it b}. However, differently from {\it panel a}, this trend is 
not reversed for satellites with high stellar mass: hosts with poor ASHs display 
systematically more surviving massive satellites ($\log M_{s,*}/M_\odot\gtrsim 9$),
despite experiencing a smaller number of HVMR AEs. 
{\it This implies that the richness of the ASH does not just follow from the number of 
HVMR AEs, but also from the survival of such massive satellites, and therefore from their infall times}. 
 
Columns in Fig.~4 correspond to the 5 quantiles $q_{* accr}$, proceeding towards 
richer ASHs to the right. 
Panels in the top row collect all 150 HAHs in each family (thin grey lines),
together with their median track, 16 and 84\% quantiles. A comparison across the different
columns shows that ordering by accreted stellar mass introduces bias: hosts with poor/rich 
ASHs experience different HAHs, despite the significant scatter. 
Hosts in the first quantile become more massive than 
average at $z\gtrsim 2$, to then follow a monotonically decreasing
track in $q_{DM}$ until recent times. Hosts in the fifth quantile
follow an orthogonal growth pattern, with an initially monotonically decreasing 
track at $z\gtrsim 1.5$, followed by a very active period at intermediate times. 
The transition between first and fifth quantiles is smooth and 
the {\it fast-slow-fast} growth pattern of hosts with poor ASHs is gradually replaced
by the {\it slow-fast-slow} pattern of hosts with rich ASHs.

The middle row of Fig.~4 shows the population of AEs that support these modes of 
growth, as a collection over the 150 HAHs. 
The timing of AEs with HVMRs ($M_s/M_h\gtrsim0.1$) is strikingly different. 
The first quintile is deficient in HVMRs AEs at intermediate times:
these hosts grow very slowly by minor mergers alone at $0.4\lesssim z\lesssim2.5$,
with a correspondingly declining $q_{DM}(z)$. At similar times, hosts
in the fifth quantile are experiencing most of their HVMR AEs, with $q_{DM}$
growing quickly. Symmetrically, the population of very recent HVMR AEs
is more numerous in hosts with poor ASHs: {\it exceedingly recent HVMR AEs provide 
an efficient way of `wasting' stellar mass, as these satellites are not destroyed by $z=0$}.
This is shown by the color-code of each AE, indicating 
the surviving fraction of stellar mass $f^*$ in $z=0$ remnants.

This analysis illustrates the best possible strategies to minimize/maximize 
$M_{\rm * accr}$ while keeping the virial mass of the host fixed. First, because 
of the steep SHMR, HVMR AEs are more efficient in contributing stars 
to the ASH (Fig.~3). Second, the timing of these AEs is equally crucial (Fig.~4):
at fixed VMR, more recent AEs contribute more stellar mass, unless they're
recent enough to survive tides. Fig.~4 shows the full spectrum between the two 
opposite strategies that make best use of these ingredients.

The mean current satellite population is the subject of the bottom row of Fig.~4: 
$M_{s,*}$ is the stellar mass of the $z\!=\!0$ remnant and color-coding of each 
event shows the surviving fraction $f^*$. 
Most HVMR AEs of hosts with poor ASHs correspond to a surviving remnant:
hosts in the first quantile feature significantly more satellites infalling 
very recently, $z_{inf}<0.5$, almost uniformly unaffected by tides,
and massive ($\log M_{s,*}/M_\odot\gtrsim 9$).  
Around hosts with rich ASHs similarly massive satellites are rare, 
and have different properties: they were accreted at earlier times and 
have already lost most of their stellar mass to the ASH. 

\begin{figure}
\centering
\includegraphics[width=\columnwidth]{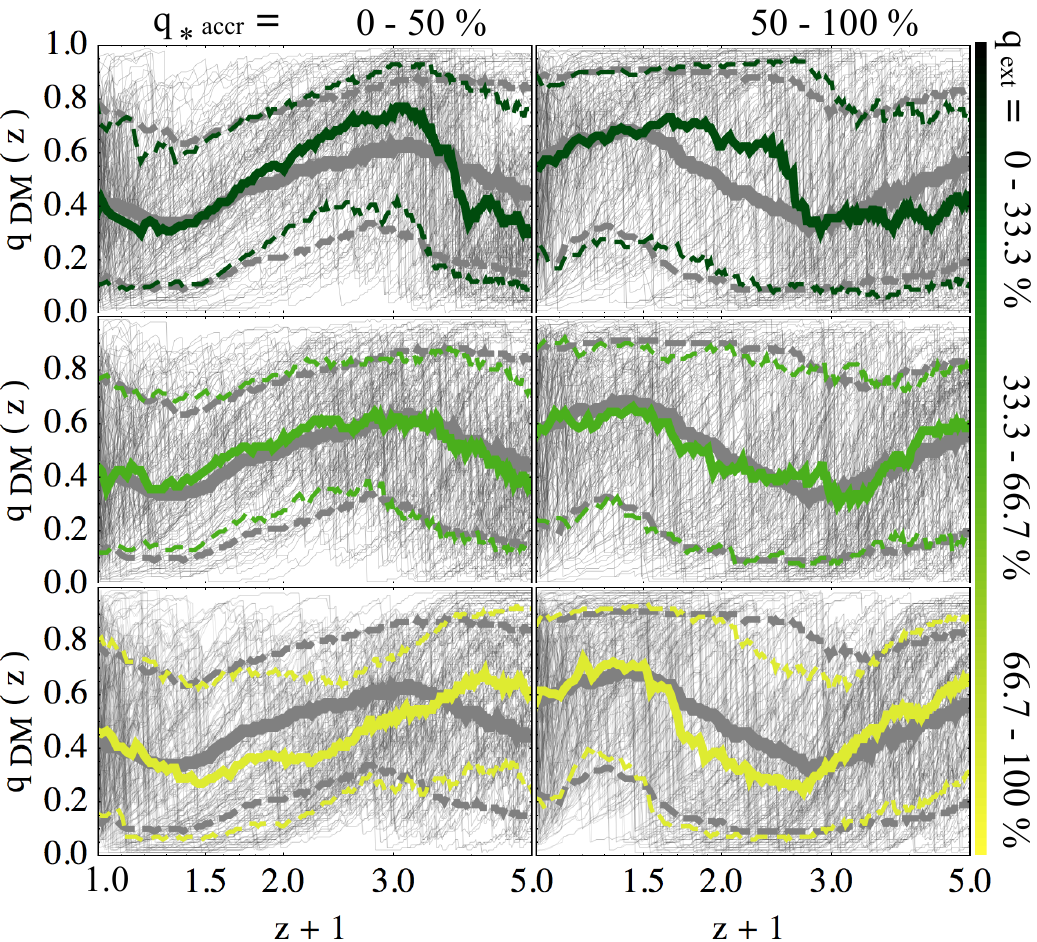}
\caption{Left (right) column: haloes with poor (rich) ASHs. Rows: 
subsequent division in terciles on the basis of the radial extension of the ASH density profile.
Thin grey lines show individual HAHs belonging to each of the 6 sub-families.
These are used to construct the colored lines, which show the 16, 50 and 84\% quantiles.
Thick grey lines are instead identical across rows, and refer to the two larger families of poor/rich 
ASHs.}
\end{figure}
%

\subsection{Disentangling HAHs by the profile of the ASH}

Fig.~5 explores on the opportunity of using the density profile of the ASH to further
constrain HAH. Columns refer to an ordering in $M_{* accr}$,
and each collect those 375 HAHs that have $q_{* accr}\leq$ or $>0.5$.
The median HAH of these two families is shown with solid grey lines in all panels 
of each column, identical across rows. 
Within these two families I introduce an additional ordering, based on the 
radial extension of the ASH profile, as quantified by the ratio 
$M_{* accr}(r>\bar r)/M_{* accr}(r\leq\bar r)$. Here, $M_{* accr}(r\leq\bar r)$ is the 
total accreted stellar mass within $\bar r$, and I have taken $\bar r=40$kpc
($25\lesssim\bar r\lesssim70$ provide very similar results). Within each column,
this ordering defines a set of terciles: hosts with 
$q_{ext}< (>)0.5$ have ASHs that are less (more) extended than the median. 

The fundamental mode of growth of all 6 families is preserved: 
independent of the `concentration' of their ASHs, hosts in the left column concur to median HAHs
(colored lines) that follow the pattern {\it fast-slow-fast}, while the orthogonal 
pattern {\it slow-fast-slow} emerges in the panels of the right column. 
Comparison with the solid grey lines shows, however, that ordering by 
$q_{ext}$ introduces additional bias, with differences in the 
intensity and timing of the different growth phases. 

For $q_{* accr}\leq0.5$, moving away from median-concentration ASHs 
implies a variation in the length of the intermediate phase of slow growth.
This is shorter in hosts with concentrated ASHs, which experience HVMR AEs at $z\gtrsim2$. 
These AEs are capable of depositing stars closer to the center of the host \citep[][]{NA16,RG16,NA17a}, resulting in a concentrated ASH. At the same time, 
they are early enough to keep $M_{* accr}$ below the median.
Hosts with equally poor but extended ASHs see this initial phase 
of fast growth pushed towards higher redshifts: they are more massive 
than the median at $z\sim4$, and therefore capable of growing through minor mergers since, 
with stellar material being deposited at comparatively larger radii.
Symmetrically, the timing of the intermediate phase of fast growth in hosts with 
$q_{* accr}>0.5$ shifts towards more recent times in proceeding from concentrated 
to extended ASHs.
Rich ASHs grow by HVMR AEs: by making them more recent 
(while still early enough to result in full satellite disruption), stellar material gets deposited 
at larger distances from the host's center, as the host gradually grows in both mass and 
size over cosmic time.

\section{Discussion and Conclusions}

This Letter shows that both global and local properties of the ASH can be 
used to constrain HA in MW-like galaxies. Hosts with 
poor/rich ASHs assemble following well defined modes of growth which minimise/maximise
the total accreted stellar mass, while keeping the final virial mass fixed.
Fundamental differences in these modes lie in the number and timing of those
rare HVMRs AEs, $M_s/M_h\gtrsim 0.1$, which dominate the stellar budget of the ASH of $L^*$ galaxies.

On average, hosts with rich ASHs experience more HVMR AEs (Fig.~3).
These take place in a phase of faster-than-average growth at 
intermediate times, $0.5\lesssim z_{inf}\lesssim1.5$,
as a best compromise between contributing the highest possible amount of stellar mass 
and allowing for sufficient time to achieve full tidal disruption by $z\!=\!0$.
In turn, at similar times, hosts with poor ASHs experience a slower-than-average growth 
sustained by minor mergers alone (Fig.~4, middle row). HVMR AEs are concentrated at very recent times, 
$z_{inf}\lesssim0.5$: these satellites are not disrupted by $z\!=\!0$ and distinctively characterise the 
surviving satellite population ($\log M_{s,*}/M_\odot\gtrsim 9$, Fig.~4, bottom row). The transition between these 
two idealised and opposite cases is smooth. Deviations in the properties of each 
contributing satellite from the mean relations
\citep[for instance: SHMR, mass-concentration-redshift relation 
and orbital parameters at infall,][]{NA17a} is mirrored in substantial scatter 
around the different growth patterns (Fig.~4, top row). 

Being a poor/rich stellar halo is in fact a temporary status: 
the massive satellites of hosts with poor ASHs will make them rich in the future. 
Poor/rich ASHs have not always being such at all redshifts, and the fundamental 
timescale of these periodic shifts is the time needed for full disruption of 
satellites with HVMRs.

The toy models used here are certainly highly simplified, and their limitations
have been analysed in recent works \citep[][]{Ba14,NA16,AC16}. 
{Additionally, at present, these models do not include 
a connection between the HAH and the properties of the host's main
stellar body: it is possible that some of the hosts in my sample would not look 
like $L_*$ galaxies, especially when approaching the edges of the ordering 
by accreted stellar mass.}
Furthermore, I have compared hosts with exactly the same virial mass and 
have ignored any {\it in-situ} contribution to the stellar halo \citep[e.g.,][]{Sh12,Do13}.

However, with its unprecedentedly large sample of explored HAHs and its purely comparative nature,
this study provides a proof of concept, and represents a first step towards the quantitative inversion of the 
connection between HAH and ASH in $L^*$ galaxies. 
{For example, this framework provides two basic predictions on the 
connection between ASH and the surviving satellite population.
\begin{itemize}
\item{Hosts with especially poor ASHs are significantly more likely to feature massive 
satellites, $\log M_{s,*}/M_\odot\gtrsim 9$; massive satellites surviving around hosts 
with rich ASHs have likely already experienced substantial stripping (Fig.~4, bottom row).}
\item{The total number of surviving satellites is smaller around hosts with rich ASHs (Fig.~3). 
When comparing our extreme quintiles in accreted stellar mass, the richness of their 
satellite populations differs by a factor of $\approx 1.5$.Below $M_{s,*}\approx10^6M_{\odot}$.
this figure can be expected to be independent of satellite mass, though it
does not account explicitly for the accretion of satellites of satellites.}
\end{itemize}}

\subsection{The MW's likely poor and concentrated ASH }

It is impossible not to note how the MW
shows hints of a poor stellar halo \citep[][and references therein]{BHG16} 
and, at the same time, it features the Magellanic Clouds: 
unusually massive satellites with very recent 
infall times \citep[e.g.,][]{GB10,JP16}. 
These are the culprits of hosts with poor ASHs for their virial mass.
Let me assume the MW indeed belongs to this class. 
Then, this framework predicts that:

\begin{itemize}
\item{Its evolution was quiet at intermediate times: undisturbed since $z\sim2$, the MW has grown slowly through 
minor mergers. This is in good agreement with previous studies \citep[e.g.,][]{FH07,AD14}. Then, HVMR AEs have characterised its very recent past, 
with the addition of Sagittarius and Magellanic Clouds, which have not yet been 
(entirely) disrupted and incorporated into the ASH.}
\item{As for those hosts featuring poor ASHs, 
the dark halo of the MW is likely to have had an early half-mass formation time. 
Therefore, it is likely to be more concentrated than average for its virial mass 
\citep[e.g.,][]{AL13}. This prediction is consistent with the results of recent 
dynamical analyses, which seem to point in similar directions \citep[][]{Ra13,SG14,Bo16}.}
\end{itemize}

Finally, in light of its sharply declining density profile \citep[e.g.,][]{BS11,AD14,CS16}
it seems reasonable to assume that the ASH of the MW is concentrated, other than poor.
If so, this framework further predicts that HVMR AEs have taken place at early times, 
$z_{inf}\gtrsim2$. This would very well agree with the known differences 
between the stellar populations of Classical dwarf Spheroidals and MW halo 
\citep[e.g.,][and references therein]{ET09,AF15}. 
The signatures of fast chemical enrichment displayed by a sizeable 
fraction of the MW halo populations would follow from these early contributions
by massive satellites with HVMRs, providing a justification to i) the high number ratio of 
blue stragglers to blue-horizontal-branch stars \citep[][]{AD15}, and ii) the distinctive 
tail of RR Lyrae stars with high amplitudes and short periods, present in the 
halo, but not in the Classical dwarf Spheroidals \citep[][]{GF15}.

A quantification of these statements is deferred to future work, 
and will be addressed by selecting those mock ASHs that are 
compatible with the observed properties of the stellar halo and 
satellite populations of MW and Andromeda, and constrain their virial mass and detailed HAHs.
Recent success in the determination of both stellar halo density profiles 
\citep[e.g.,][]{AMe16,Ha16} and satellite populations \citep[][]{DC16,ET16}
of nearby galaxies will soon enable similar studies on a sample of
external galaxies.

\section*{Acknowledgements}

I am indebted with Antonela Monachesi for her useful comments
on an early version of this manuscript. It is a pleasure to thank the anonymous
referee for a constructive report.



\end{document}